# Using Large Language Models for Emotional Support of Bulgarian Users: A Survey

Melania Berbatova
Faculty of Mathematics and Informatics - Sofia University "St. Kliment Ohridski", Bulgaria
msberbatova@fmi.uni-sofia.bg

*Abstract:* The use of large language models (LLMs) for psychological and emotional support (ES) has rapidly evolved, becoming the most widely used application of generative artificial intelligence among consumers by 2025. This paper presents the results of an anonymous survey of 100 Bulgarian users, primarily high school, university, and doctoral students, to explore their attitudes toward and usage of chatbots for emotional support. Findings indicate that approximately one-half of the surveyed population utilizes chatbots for ES, with ChatGPT being the most dominant platform. Users primarily seek support for coping with stress in interpersonal relationships and work or study-related environments. While 71% of users perceive the technology as effective, non-users remain sceptical. Despite the growing adoption, significant concerns persist regarding data security, technology reliability, and the tendency of chatbots to provide excessive affirmation.

## 1. Introduction

The field of using large language models for psychological and emotional help has rapidly evolved in recent months, with a large volume of published scientific and journalistic studies. According to a study published in Harvard Business Review [1], emotional support is the most widely used application of generative artificial intelligence among consumers in 2025. In this paper, we present the results of a survey of 100 Bulgarians on their use of large language models for emotional and psychological support.

### 1.1. Emotional support

Emotional support (ES) in theory is a communication style which aims at reducing individuals' emotional distress and helping them understand and work through the challenges that they face. Research shows that recipients of sensitive emotional support feel better and cope with problems more effectively and may even be healthier. Contrary, recipients of insensitive emotional support may feel worse than ever, and even suffer from stress-related illness [2].

According to Lehman and Hampill [7], the characteristics of effective emotional support include:

- expressing love, concern, or understanding;
- encouragement;
- listening;
- praising abilities;
- involvement in social activities;
- emotional presence ("being there").

In essence, emotional support, unlike psychotherapy, does not involve long-term problem solving, but rather short-term coping with emotional stress. However, ES can help individuals develop coping mechanisms for stressful situations, thus leading to the development of a more resilient psyche in those receiving support.

### 1.2. Challenges of traditional methods

Today, traditional therapy faces several challenges. Therapy sessions are expensive and often hard to organize, and in many cultures, there is a social stigma around participating in them [3]. On the other hand, there is a global shortage of mental health workers – a survey conducted by the World Health Organizations (WHO) in 2022 [4] showed that on average, there are 13 mental health workers for average 100,000 people.

Furthermore, events such as pandemics, wars, and natural disasters lead to circumstances in which millions of people simultaneously need emotional and mental support. It is estimated that only the Covid-19 pandemic triggered an additional 53 million cases of depression and 76 million cases of anxiety globally [5]. In such situations it is impossible for everyone affected to get personal support and automatic systems may be of great help in dealing with massive emotional distress.

### 1.3. Risks and benefits of AI assistants

In the context of emotional support, there are many benefits of using personal AI assistants. First, they can be available anytime, day or night. They are significantly cheaper than traditional therapy, if not completely free. It is also important to note that studies show that some people feel more comfortable confessing their feelings to a robot rather than a person [6].

Despite their indisputable benefits, the use of AI for emotional support poses some significant risks. Some of the most challenging and dangerous risks noted in literature [8] include:

- **Misinformation:** Fluency is frequently conflated with factuality in LLMs, which generate credible answers that may not align with real-life evidence. Hallucinated responses, especially when presented in a confident tone, can dangerously mislead users [9].

- **Persuasion and manipulation:** LLM-based chatbots can generate highly persuasive messages, raising concerns in mental health contexts where users are frequently in a vulnerable state.

- **Over-reliance:** Users and clinicians may overtrust such AI assistants, treating their outputs as undoubtedly truthful and reliable. Fluency and confidence in responses can mislead users into accepting poor advice [9]. Repeated use of chatbots may also lead to emotional dependency and reduced engagement with human care [10].

## 2. Survey of Bulgarian users

In order to gain a better insight into the attitudes of the Bulgarian population towards such a technology, we conducted an anonymous survey on the usage of chatbots for emotional support among Bulgarian-speaking internet users. We obtained 100 survey answers from respondents from different age groups and sociological and educational backgrounds.

### 2.1 Demographics

The survey was completed by 100 people online, mainly high school, university and doctoral students from various disciplines, as well as a community of young professionals in Bulgaria. Therefore, while the distribution of the surveyed population is not representative of the overall Bulgarian population, we were able to obtain answers from participants of various ages and educational levels.

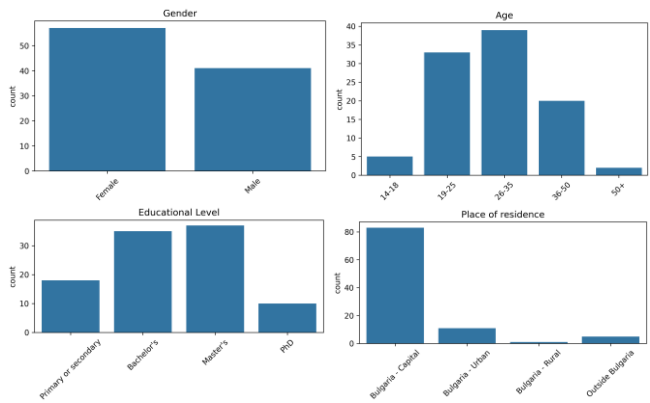

*Fig. 1 Demographics of the surveyed population.*

### 2.2 Usage of chatbots for ES

Roughly one-half of the surveyed population (Figure 2) uses chatbots for emotional support. As shown in Figure 3, the most used chatbot for emotional support is, without competition, ChatGPT. The survey reveals only isolated instances of usage of other language models, such as Gemini, Claude, Grok3, and Gemma. When it comes to preferred language for communication, the majority of users state that they mix Bulgarian and English, followed by English-only usage and Bulgarian-only usage.

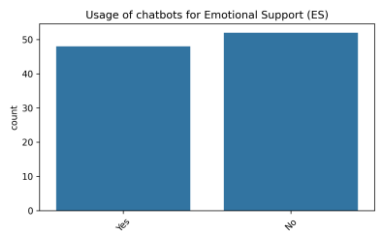

*Fig. 2 Usage of chatbots for Emotional Support (ES)*

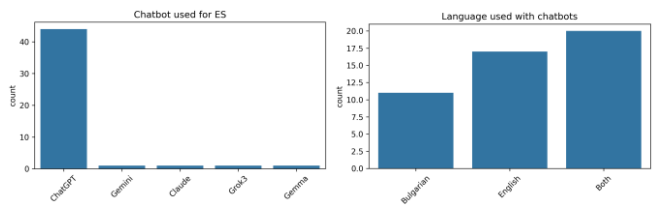

*Fig. 3 Characteristics of usage*

### 2.3 Purposes, concerns and perceived effectiveness

The most frequently stated purpose of using chatbots for ES is coping with stress in interpersonal relationships (37 respondents), followed by stress related to education or work (27 respondents). 13 people indicate that they use chatbots to cope with the symptoms of more prolonged mental health issues, such as depression, bipolar disorder, and others. 3 people add other purposes, such as psychoeducation or personal and spiritual development. Figure 4 illustrates the distribution.

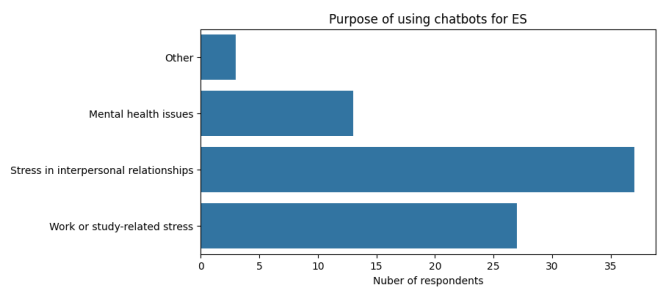

*Fig. 4 Purpose of using chatbots for ES*

The most frequently stated concerns (Figure 5) are data confidentiality and technology reliability (with 35% of users citing these as their concerns), followed by concerns about long-term negative effects on individuals and interpersonal relationships (21%). Twenty-nine percent of users say they have no concerns about the technology. The most frequently listed reason for concern outside of the predefined possible answers is the tendency of chatbots to always agree and affirm the user's point of view (8%).

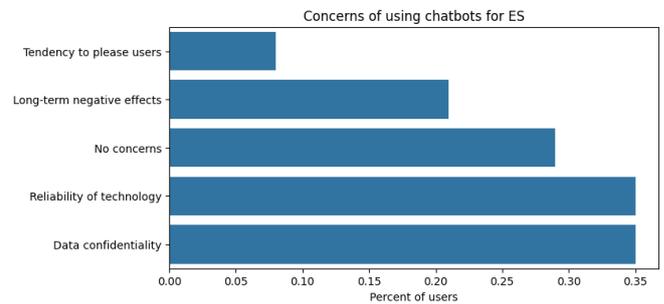

*Fig. 5 Concerns of using chatbots for ES*

As shown in Figure 6, 71% of users believe that such technology is effective, 19% cannot decide, and 10% do not believe in its effectiveness. By comparison, among respondents who do not use the technology, 55% believe it is not effective, only 6% believe it is effective, and 37% cannot decide.

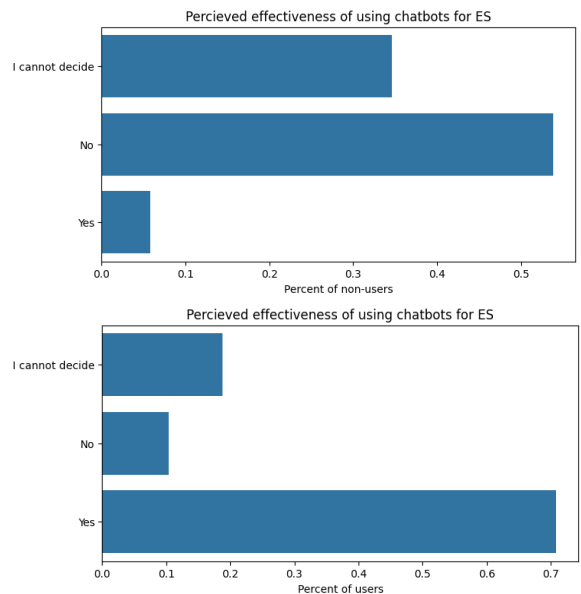

*Fig. 6 Perceived effectiveness among users and non-users*

### 2.4 Demographics by usage

Figure 7 shows the joint demographic distributions in response to the question "Do you use AI-based chatbots as a means of providing emotional support?".

The statistical analysis of the distributions showed the following correlations:

• **Gender**: no statistically significant correlation ($p = 0.344$).

• **Age**: a tendency towards a correlation, but not statistically significant ($p = 0.069$).

• **Level of education**: no significant correlation ($p = 0.423$).

• **Place of residence:** no significant correlation ($p = 0.443$).

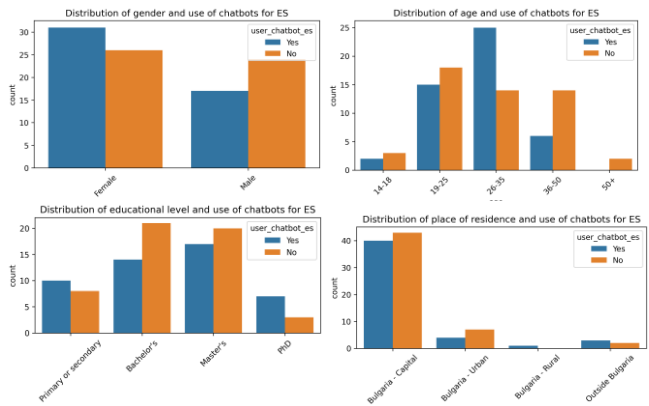

*Fig. 7 Demographics by usage*

Overall, no significant correlation between demographic characteristics and usage of AI-based chatbots for emotional support was found, meaning that both adopters and skeptics of such a technology occur across different age ranges, genders, educational levels and places of residence. The "age" category was the only one in which a tendency for correlation was found, showing that such use of the technology is mostly adopted by the 26–35 years' old (people born 1990–1999, often informally referred to as "Zillenials" and "90's kids").

## 3. Qualitative research: Perspectives of users and specialists

In order to further get insights from the quantitative data from the survey, qualitative research was conducted through semi-structured interviews with users (non-specialists) and experts in the field of mental health. The results reveal a nuanced understanding of the role of AI as a tool for emotional support.

### 3.1. The chatbot as a tool for self-reflection and "venting"

The interviewed users identified several key functions of the LLM as an emotional supporter:

**Guided journaling:** Users describe the interaction as a form of "guided journaling" that helps structure thoughts and needs. *"We often say, 'Sit down and write it out' or 'Keep a journal'. Here, however, there is something a little different – because you get a response. Often, this response is filled with empathy and can encourage you to continue sharing more and more."*

**Safe sharing**: The chatbot is perceived as a "friend" to whom you can "vent" without the risk of embarrassing yourself or other people in front of loved ones or family.

**Rationalization and cognitive reframing:** The models help change your perspective (reframing) by asking clarifying questions. A respondent shares: *"It helps you change your perspective if you ask good questions. In my opinion, the benefit depends on how well you use these models in general. As questions, I ask: 'Ask me several clarifying questions to help me decide what to do.' or 'Is there anything I am overlooking in this situation?'"*

**No social pressure**: For some users, AI is preferable to a psychologist because it eliminates the need to carefully consider their words and allows them to dive straight into the desired topic, without the need to agree with the psycologist's point of view.

**Encouragement and validation:** An interesting case from the qualitative interviews involves a medical professional using AI to manage postpartum stress. Despite her professional background, she noted a distinct need: *"I need to communicate with someone who is as knowledgeable as I am and can give me adequate advice."* She shares that her primary motive was not information retrieval, but emotional support and the need for a non-judgmental, yet knowledgeable "listener" to validate her maternal decisions – *"And even though I ask about medical conditions I have read about, I realize the responsibility of the decisions I make as a mother, and I need someone to encourage me."*

### 3.2. The sceptics' point of view

The majority of non-users (55%) believe that technology is ineffective for emotional support purposes. Skeptics believe that human connection and personal relationships in the realm of mental health and emotional well-being cannot be replaced by technology. They prefer to receive understanding and attention from a real person (psychologist or psychotherapist) rather than talking to a chatbot.

Some users prefer to cope on their own or talk to friends in difficult moments – *"I see no point in using artificial intelligence in the near future."*

### 3.2. The professional perspective

Interviews with specialists highlight the critical difference between "emotional relief" and "psychological change". The experts emphasis on the lack of a biological basis for empathy: AI does not have "mirror neurons" and cannot establish real live contact. While therapists use emotional reactions (transference) to analyze clients' behavior patterns, AI is not emotionally intelligent and cannot sense repressed emotions. Further, experts warn on the risk of superficial validation, as chatbots often gives the user exactly what they want and ask for, rather than working with their psychological defenses.

## 4. Discussion

The findings of this research suggest a nuanced landscape of AI adoption for emotional support among Bulgarian users, characterized by a preference for multilingual support. The linguistic switching between English and Bulgarian may be due to the need of expressing both culturally specific emotional states and terminology, typical for the English language.

A central demographic trend identified in this study is the high adoption rate among individuals born between 1990 and 1999, often referred to as "Zillenials" [11]. Although the statistical correlation for age showed only a tendency rather than full significance (p=0.069), this group emerges as the primary user base. The group is characterized as an intersection of high digital literacy and significant life stressors, such as early-career pressures, early parenthood and evolving interpersonal relationships. For these users, the chatbot often serves as a "guided journal" or a "rational friend" that provides a non-judgmental space to "blow off steam" without the social risks.

However, the perceived effectiveness of these models – reported at 71% among users – must be weighed against the risk of "social sycophancy" [12], or the tendency of LLMs to provide excessive affirmation. This "flattery" concern was explicitly noted by 8% of respondents who worried that chatbots consistently take the user's side. As suggested in the qualitative interviews, this tendency can be counterproductive, as the AI may simply validate the user's current state rather than challenging their views in the discussed situation. This poses the risk of over-reliance: users may stop seeking human intervention if the AI always agrees with them. While this kind technology is excellent at providing immediate support, it lacks genuine empathy and the ability to work with complex phenomena like transference and countertransference [13].

## 5. Conclusion

The results of the presented study suggest that there is no consensus among the Bulgarian users on the extent to which chatbots based on LLMs are a useful technology for providing emotional support or not. Based on the conducted survey, the proportion of people who use them for this purpose is approximately equal to those who do not. Among the stated purposes for use, coping with stress in interpersonal relationships stands out, which certainly affects the way people communicate with each other – now and in the near future. No statistically significant correlation was found between attitudes and use of the technology and demographic characteristics such as gender, level of education, age, and place of residence. The only notable exception is the group of people born between 1990 and 1999, which emerge as the predominant adopter group of the technology within this sample. This trend demonstrates a shift towards searching proactive, digital and conversational self-help, especially on the topics of parenthood, family life, and career development.

The biggest concerns stated by the surveyees are related to the reliability of the technology and data privacy, but tendency of models to provide excessive affirmation or "flattery" also stands out. This tendency can create emotional echo chambers that prevents psychological growth and encourage dangerous behavior or thought patterns. A primary recommendation for developers of such technology is a movement toward "constructive challenging" – designing models that can identify and gently confront a user's harmful thought patterns rather than simply validating them.

In summary, while LLMs offer a solution to some of the shortcomings of traditional therapy by providing an always-available "guided journal", they must be viewed as a supplemental technology for resilience rather than a replacement for the genuine human empathy and the deep psychological work inherent in human-to-human therapy.

## *Acknowledgements*

The author gratefully acknowledge the support provided by the project UNITe BG16RFPR002-1.014-0004 funded by PRIDST.